\documentclass{article}
\usepackage[utf8]{inputenc}
\usepackage{graphicx}
\usepackage{amsmath}
\usepackage{booktabs}
\usepackage{hyperref}
\usepackage[capitalise]{cleveref}
\usepackage[a4paper, total={6in, 8in}]{geometry}
\usepackage{authblk}
\usepackage{amssymb}

\title{DeepInverse: a Python package for solving imaging inverse problems with deep learning}

\author[1,†\footnote{† denotes equal contribution.}]{Julián Tachella}
\author[2,†]{Matthieu Terris}
\author[3,†]{Samuel Hurault}
\author[4,†]{Andrew Wang}
\author[5]{Dongdong Chen}
\author[8]{Minh-Hai Nguyen}
\author[14]{Maxime Song}
\author[4,5]{Thomas Davies}
\author[1]{Leo Davy}
\author[7]{Jonathan Dong}
\author[9]{Paul Escande}
\author[15]{Johannes Hertrich}
\author[7]{Zhiyuan Hu}
\author[13]{Tobías I. Liaudat}
\author[6]{Nils Laurent}
\author[12]{Brett Levac}
\author[10]{Mathurin Massias}
\author[2]{Thomas Moreau}
\author[11]{Thibaut Modrzyk}
\author[6]{Brayan Monroy}
\author[16]{Sebastian Neumayer}
\author[1]{Jérémy Scanvic}
\author[8]{Florian Sarron}
\author[1]{Victor Sechaud}
\author[17]{Georg Schramm}
\author[1]{Romain Vo}
\author[8]{Pierre Weiss}

\affil[1]{CNRS, ENS de Lyon, Univ Lyon, Lyon, France}
\affil[2]{Université Paris-Saclay, Inria, CEA, Palaiseau, France}
\affil[3]{CNRS, ENS Paris, PSL, Paris, France}
\affil[4]{University of Edinburgh, Edinburgh, UK}
\affil[5]{Heriot-Watt University, Edinburgh, UK}
\affil[6]{Universidad Industrial de Santander, Bucaramanga, Colombia}
\affil[7]{EPFL, Lausanne, Switzerland}
\affil[8]{IRIT, CBI, CNRS, Université de Toulouse, Toulouse, France}
\affil[9]{IMT, CNRS, Université de Toulouse, Toulouse, France}
\affil[10]{Inria, ENS de Lyon, Univ Lyon, Lyon, France}
\affil[11]{INSA de Lyon, Lyon, France}
\affil[12]{University of Texas at Austin, Austin, USA}
\affil[13]{IRFU, CEA, Université Paris-Saclay, Gif-sur-Yvette, France}
\affil[14]{CNRS UAR 851, Université Paris-Saclay Orsay, France}
\affil[15]{Université Paris Dauphine - PSL, Paris, France}
\affil[16]{Chemnitz University of Technology, Chemnitz, Germany}
\affil[17]{Department of Imaging and Pathology, KU Leuven, Leuven, Belgium}

\newcommand{\code}[1]{\texttt{#1}}
\usepackage{amsmath}

\DeclareMathOperator*{\argmin}{arg\,min}

\newcommand{\double}[2]{\multicolumn{1}{l}{\begin{tabular}[t]{@{}l@{}}#1\\ #2\end{tabular}}}

\date{May 15, 2025}
\begin{document}

\maketitle

\begin{abstract}
DeepInverse is an open-source PyTorch-based library for solving imaging inverse problems. The library covers all crucial steps in image reconstruction from the efficient implementation of forward operators (e.g., optics, MRI, tomography), to the definition and resolution of variational problems and the design and training of advanced neural network architectures. In this paper, we describe the main functionality of the library and discuss the main design choices.
\end{abstract}

\section{Statement of Need}

Deep neural networks have become ubiquitous in various imaging inverse problems, from computational photography to astronomical and medical imaging. Despite the ever-increasing research effort in the field, most learning-based algorithms are built from scratch, are hard to generalize beyond their specific training setting, and the reported results are often hard to reproduce. DeepInverse overcomes these limitations by providing a unified framework for defining imaging operators and solvers. It leverages the popular PyTorch deep learning library \cite{paszke2019pytorch}, making most modules compatible with auto-differentiation.
The target audience of this library is both researchers in inverse problems (experts in optimization, machine learning, etc.) and practitioners (biologists, physicists, etc.). DeepInverse has the following objectives:

\begin{itemize}
    \item Accelerate research by enabling efficient testing, deployment, and transfer of new ideas across imaging domains;
    \item Enlarge the adoption of deep learning in inverse problems by lowering the entrance bar to new researchers and practitioners;
    \item Enhance research reproducibility via a common framework for imaging operators, reconstruction methods, datasets, and metrics for inverse problems.
\end{itemize}

While other Python computational imaging libraries exist, to the best of our knowledge, DeepInverse is the only one with a strong focus on learning-based methods, providing a larger set of realistic imaging operators.
SCICO \cite{balke2022scico} and Pyxu \cite{simeoni2022pyxu} are Python libraries whose main focus is variational optimization and/or plug-and-play reconstruction methods. These libraries do not provide specific tools for training reconstruction models such as trainers and custom loss functions, and do not cover non-optimization-based solvers including diffusion methods, adversarial methods, or unrolling networks.
CUQIpy \cite{riis2024cuqipy} is a library focusing on Bayesian uncertainty quantification methods for inverse problems.
TIGRE \cite{biguri2025tigre}, ODL \cite{adler2018odl}, and CIL \cite{jorgensen2021core} mostly focus on computed tomography and do not cover deep learning pipelines for inverse solvers. 
There are also multiple libraries focusing on specific inverse problems: ASTRA \cite{van2016astra} and the related pytomography \cite{polson2025pytomography} define advanced tomography operators, sigpy \cite{ong2019sigpy} provides magnetic resonance imaging (MRI) operators without deep learning, and PyLops \cite{ravasi2019pylops} provides a linear operator class and many built-in linear operators. These operator-specific libraries can be used together with DeepInverse as long as they are compatible with PyTorch (for example, we provide a wrapper for ASTRA). Advanced libraries for inverse problems also exist in other programming languages such as MATLAB, including GlobalBioIm \cite{soubies2019pocket} or IR Tools \cite{gazzola2019ir}, but they are restricted to handcrafted reconstruction methods without automatic differentiation. 

\begin{figure}[t]
    \centering
    \includegraphics[width=\linewidth]{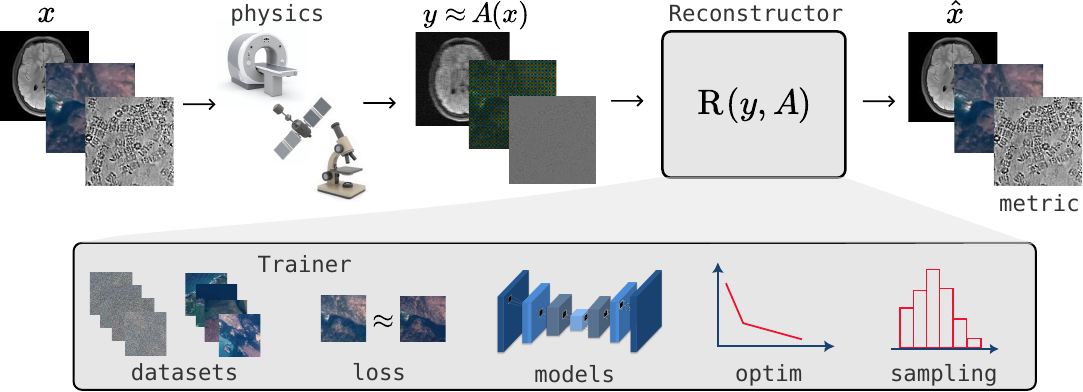}
    \caption{Schematic of the main modules of the library.}
    \label{fig:schematic}
\end{figure}

\section{Inverse Problems}

Imaging inverse problems can be expressed as 
\begin{equation} \label{eq:forward}
y = N_{\sigma}(A_{\xi}(x))
\end{equation}
where $x\in\mathcal{X}$ is an image, $y\in\mathcal{Y}$ are the measurements, $A_{\xi}\colon\mathcal{X}\mapsto\mathcal{Y}$ is a
deterministic (linear or non-linear) operator capturing the physics of the acquisition and
$N_{\sigma}\colon\mathcal{Y}\mapsto \mathcal{Y}$ is a mapping that characterizes the noise affecting the 
measurements parameterized by $\sigma$ (e.g. the noise level and/or gain). The forward operation is simply written in the library as \code{x = physics(y, **params)} where \code{params} is a dictionary with
optional parameters $\xi$. This framework unifies the wide variety of forward operators across various domains. Most forward operators in \code{deepinv} are matrix-free, scaling gracefully to large image sizes.
The library provides high-level operator definitions which are associated with specific imaging applications 
(MRI, computed tomography, radioastronomy, etc.), and allows users to perform operator algebra,
like summing, concatenating or stacking operators. \code{deepinv} comes with multiple useful tools for handling
linear operators, such as adjoint, pseudoinverses, and proximal operators (leveraging the singular value decomposition when possible), matrix-free linear solvers \cite{hestenes1952methods,paige1975solution,van1992bi},
and operator norm and condition number estimators \cite{paige1982lsqr}. Many common noise distributions are included in the library
such as Gaussian, Poisson, mixed Poisson-Gaussian, uniform and gamma noise. Table \ref{tab:operators} summarizes the available forward operators
at the time of writing, which are constantly being expanded and improved upon by the community.


\renewcommand{\arraystretch}{1.2} 

\begin{table}[t]
\centering
\label{tab:operators}
\begin{tabular}{lll}
\toprule
\textbf{Family} & \textbf{Operators} $A$ & \textbf{Generators} $\xi$ \\
\midrule
Pixelwise &\double{Denoising, inpainting,}{demosaicing, decolorize}  &\double{Mask generators,}{noise level generators}  \\
Blur \& Super-Resolution & \double{Blur, downsampling,}{space-varying blur} & \double{Motion, Gaussian,}{diffraction blurs} \\
MRI & \double{Single/multi-coil,}{dynamic/sequential (3D support)} & Gaussian, random masks \\
Tomography & 2D/3D parallel beam, fanbeam, conebeam & - \\
Remote Sensing & Pansharpening, hyperspectral unmixing & - \\
Compressive & Compressed sensing, single-pixel camera & - \\
Radio Interferometry & Monochromatic intensity imaging & - \\
Single-photon Lidar & TCSPC lidar & - \\
Dehazing & Parametric haze model & - \\
Phase Retrieval & Random operators, ptychography & Probe generation \\
\bottomrule
\end{tabular}
\caption{Forward operators implemented in the library (v0.3.0).}
\end{table}

\subsection{Operator Parameterization}

Most physics operators are parameterized by a vector $\xi$, which has a different meaning depending on the context.
For instance, it can represent the projection angles in tomography, the blur kernel in image deblurring, the acceleration masks in MRI, etc.
Integrating this parameter allows for advanced computational imaging problems, including calibration of the system (measuring $\xi$ from $y$),
blind inverse problems (recovering $\xi$ and $x$ from $y$) \cite{debarnot2024deep,chung2023parallel}, co-design (optimizing $\xi$ and possibly 
the reconstruction algorithm jointly) \cite{lazarus2019sparkling,nehme2020deepstorm3d}, robust neural network training \cite{gossard2024training,terris2023meta,terris2025ram}. To the best of our knowledge, this feature is distinctive and becoming essential in recent advances in image reconstruction.

\section{Reconstruction Methods}

The library provides multiple solvers that depend on the forward operator $A_{\xi}$ and the noise distribution via $\sigma$. Our framework unifies the wide variety of solvers that are commonly used in the current literature:
\begin{equation} \label{eq:solver}
\hat{x} = \operatorname{R}_{\theta}(y, A_{\xi}, \sigma)
\end{equation}
where $\operatorname{R}_{\theta}$ is a reconstruction network/algorithm with (optional) trainable parameters $\theta$ (see Section \ref{sec: training}).
In \code{deepinv} code, a reconstructor is simply evaluated as \code{x\_hat = model(y, physics)}.
The library covers a wide variety of approaches for building $\operatorname{R}_{\theta}$, which can be roughly divided into
optimization-based methods, sampling-based methods, and non-iterative methods. 

\subsection{Optimization-Based Methods}

These methods consist of solving an optimization problem  \cite{chambolle2016introduction}
\begin{equation} \label{eq:var}
\operatorname{R}_{\theta}(y, A_{\xi}, \sigma) \in \argmin_{x} \, f_{\sigma}(y,A_{\xi}(x)) + g(x)
\end{equation}
where $f_{\sigma}\colon\mathcal{Y} \times \mathcal{Y} \mapsto \mathbb{R}$ is the data fidelity term, which can incorporate knowledge about the noise parameters $\sigma$, and $g\colon\mathcal{X}\mapsto\mathbb{R}$ is a regularizer that promotes plausible reconstructions.
The \code{optim} module includes classical fidelity terms (e.g., $\ell_1$, $\ell_2$, Poisson log-likelihood) and offers a wide range of regularization priors:

\paragraph{Explicit Priors} The library implements several traditional regularizers, such as sparsity \cite{candes2008introduction}, total variation \cite{rudin1992nonlinear}, wavelets \cite{stephane1999wavelet}, patch-based likelihoods \cite{zoran2011learning}, and mixed-norm regularizations \cite{kowalski2009sparse}.
We also provide learned regularizers such as EPLL \cite{zoran2011epll} and PatchNR \cite{altekruger2023patchnr}.

\paragraph{Denoising Plug-and-Play Priors} Plug-and-Play (PnP) methods replace the proximal operator \cite{venkatakrishnan2013plug} or the gradient \cite{romano2017little} of the regularizer $g$ with a pretrained denoiser, often based on deep learning. The library provides access to widely used pretrained denoisers $\operatorname{D}_{\sigma}$, many of them trained on multiple noise levels $\sigma$, including DnCNN \cite{zhang2017beyond}, DRUNet \cite{zhang2021plug}, and recent diffusion-based denoisers such as DDPM \cite{ho2020denoising} and NCSN \cite{song2020score}.


\paragraph{Optimization Algorithms} The library contains several classical algorithms \cite{dossal2024optimizationorderalgorithms} for minimizing the sum of two functions, including proximal gradient descent, FISTA, ADMM, Douglas-Rachford Splitting, and primal-dual methods.

\paragraph{Unfolded Networks and Deep Equilibrium Models} Unfolded networks \cite{gregor2010learning} are obtained by unrolling a fixed number of iterations of an optimization algorithm and training the parameters end-to-end, including both optimization hyperparameters and deep regularization priors. Standard unfolded methods train via backpropagation through the optimization algorithm, while deep equilibrium methods \cite{bai2019deep} implicitly differentiate the fixed point of the algorithm.

\subsection{Sampling-Based Methods}
Reconstruction methods can also be defined via ordinary or stochastic differential equations, generally as a Markov chain defined by
\begin{equation}
x_{t+1} \sim p(x_{t+1}|x_t, y, \operatorname{R}_{\theta}, A_{\xi}, \sigma) \text{ for } t=0,\dots,T-1
\end{equation} 
such that $x_{T}$ is approximately sampled from the posterior $p(x|y)$, and $\operatorname{R}_{\theta}$ is a (potentially learned) denoiser.
Sampling multiple plausible reconstructions enables uncertainty estimates by computing statistics across the samples.
The sampling module provides implementations of the following methods:

\paragraph{Diffusion Models} In a similar fashion to PnP methods, diffusion models \cite{chung2022diffusion,kawar2022denoising,zhu2023denoising} incorporate prior information via a pretrained denoiser, however, they are linked to a stochastic differential equation (SDE) or an ordinary differential equation (ODE), instead of the optimization of \eqref{eq:var}.

\paragraph{Langevin-Type Algorithms} The library provides popular high-dimensional Markov Chain Monte Carlo (MCMC) methods such as unadjusted Langevin algorithm and some of its variants \cite{laumont2022bayesian,pereyra2020skrock}, which
define a Markov chain with stationary distribution close to the posterior distribution $p(x|y) \propto \exp(f_{\sigma}(y, A_{\xi}(x)) + g(x))$.

\subsection{Non-Iterative Methods}

Non-iterative methods are part of the \code{models} module, and include artifact removal, unconditional and conditional generative networks, and foundation models.

\paragraph{Artifact Removal} The simplest way of incorporating the forward operator into a network architecture is to backproject the measurements to the image domain and apply a denoiser (image-to-image) architecture $\operatorname{D}_{\sigma}$ such as a UNet \cite{jin2017deep}.
These architectures can be thus written as

\begin{equation}
\operatorname{R}_{\theta}(y, A_{\xi}, \sigma) = \operatorname{D}_{\sigma}(A_{\xi}^{\top}y),
\end{equation}
where the back-projection can be replaced by any pseudoinverse of $A_{\xi}$. 

\paragraph{Unconditional Generative Networks} Generative models exist in unconditional or conditional forms. Unconditional methods \cite{bora2017compressed,bora2018ambientgan} leverage
a pretrained generator $\operatorname{G}_{\theta}(z)\colon\mathcal{Z}\mapsto \mathcal{X}$ where $z\in\mathcal{Z}$ is a latent code to solve an inverse problem via
\begin{equation} \label{eq:cond}
\operatorname{R}_{\theta}(y, A_{\xi}, \sigma) = \operatorname{G}_{\theta}(\hat{z}) \text{ with } \hat{z} \in \operatorname{argmin}_{z} f_{\sigma}(y,A_{\xi}(\operatorname{G}_{\theta}(z)))
\end{equation}
The deep image prior \cite{ulyanov2018deep} is a specific case of unconditional models that uses an untrained generator $\operatorname{G}_{\theta}$, leveraging the inductive bias of a specific autoencoder architecture. 

\paragraph{Conditional Generative Networks} Conditional generative adversarial networks \cite{isola2017image,bendel2023gan} use adversarial training to learn a network $\operatorname{R}_{\theta}(y, z, A_{\xi}, \sigma)$ which provides a set of reconstructions by sampling different latent codes $z\in\mathcal{Z}$. 

\paragraph{Foundation Models} Foundation models are end-to-end architectures that incorporate knowledge of $A_{\xi}$ and $\sigma$ and are trained to reconstruct images across a wide variety of forward operators $A_{\xi}$ and noise distributions $N_{\sigma}$ \cite{terris2025ram}.
These models often obtain good performance in new tasks without retraining, and can also be finetuned to specific inverse problems or datasets using measurement data alone.

\begin{table}[t]
\centering
\label{tab:methods}
\begin{tabular}{lllll}
\toprule
\textbf{Method Family} & \textbf{Description} & \textbf{Training} & \textbf{Iterative} & \textbf{Sampling} \\
\midrule
Artifact Removal & Applies network to pseudoinverse & Yes & No & No \\
Variational & Optimization with priors & Hyperparams & Yes & No \\
PnP & Pretrained denoisers as priors & Hyperparams & Yes & No \\
Unfolded & Unrolled optimization & Yes & Only DEQ & No \\
Diffusion & Pretrained denoisers in ODE/SDE & Hyperparams & Yes & Yes \\
GANs & Model via generator & Yes & Yes & Depends \\
Foundation models & Multi-problem models & Finetune & No & No \\
\bottomrule
\end{tabular}
\caption{Reconstruction methods implemented in the library (v0.3.0).}
\end{table}

\section{Training} \label{sec: training}

Training can be performed using the \code{Trainer} class, which is a high-level interface for training reconstruction networks. \code{Trainer} handles data ingestion, the training loop, logging, and checkpointing. 

\subsection{Losses}

The library also provides the `loss` module with losses for training $\operatorname{R}_{\theta}$ which are specially designed for inverse problems, acting as a framework that unifies loss functions that are widely used in various inverse problems across domains. Loss functions are defined as

\begin{equation} \label{eq:loss}
l = \mathcal{L}\left(\hat{x}, x, y, A_{\xi}, \operatorname{R}_{\theta}\right)
\end{equation}
with $\hat{x}=R_{\theta}(y,A_{\xi},\sigma)$ being the network prediction, 
and written in \code{deepinv} as \code{l = loss(x\_hat, x, y, physics, model)}, where
some inputs might be optional (e.g., $x$ is not needed for self-supervised losses).

\paragraph{Supervised Losses} Supervised learning can be done using a dataset of ground-truth and measurements pairs $\{(x_i,y_i)\}_{i=1}^{N}$ by applying a metric to compute the distance between $x$ and $\hat{x}$.
If the forward model is known, measurements are typically generated directly during training from a dataset of ground-truth references $\{x_i\}_{i=1}^{N}$.

\paragraph{Self-Supervised Losses} Self-supervised losses rely on measurement data only $\{y_i\}_{i=1}^{N}$. We provide implementations of state-of-the-art losses from the literature \cite{wang2025benchmarking}. These can be roughly classified into three classes:

The first class consists of splitting losses \cite{batson2019noise2self}, with operator-specific solutions for denoising \cite{krull2019noise2void,huang2021neighbor2neighbor} and MRI \cite{yaman2020self,eldeniz2021phase2phase,liu2020rare}.
The second class is Stein's Unbiased Risk Estimate (SURE) and related losses: we provide variants of SURE for Gaussian, Poisson, and Poisson-Gaussian noise respectively, which can also be used without knowledge of the noise levels \cite{tachella2025unsure}. The library includes the closely related Recorrupted2Recorrupted \cite{pang2021recorrupted} which handles Gaussian, Poisson, and Gamma noise distributions \cite{monroy2025gr2r}.
The third class corresponds to nullspace losses, exploiting invariance of the image distribution to transformations \cite{chen2021equivariant} or access to multiple forward operators \cite{tachella2022unsupervised}. 

The library provides image transforms $\tilde{x}=T_g x$ in the \code{transform} module where $g\in G$ parametrizes a transformation group, including a group-theoretic model for geometric transforms covering linear (Euclidean) and non-linear (projective, diffeomorphism) transforms \cite{wang2024perspective}. These can be used for data augmentation, equivariant imaging \cite{chen2021equivariant}, and equivariant networks.

\paragraph{Network Regularization Losses}  Network regularization losses which try to enforce some regularity condition on $\operatorname{R}_{\theta}$, such as having an upper bounded Lipschitz constant or similarly being firmly non-expansive \cite{pesquet2021learning}.

\paragraph{Adversarial Losses}  Adversarial losses are used to train conditional or unconditional generative networks. The library provides 
both supervised \cite{bora2017compressed} and self-supervised (i.e., no ground-truth) \cite{bora2018ambientgan} adversarial losses.

\section{Datasets}

The library provides a common framework for defining and simulating datasets for image reconstruction. Datasets return ground-truth and measurements pairs $\{(x_i,y_i)\}_{i=1}^{N}$, and may also return physics parameters $\xi_i$. Given a dataset of reference images $\{x_i\}_{i=1}^{N}$, the library can be used to generate and save a simulated paired dataset to encourage reproducibility. The library also provides interfaces to some popular datasets to facilitate research in specific application domains: 

\begin{itemize}
    \item Natural images: Div2K \cite{agustsson2017ntire}, Urban100 \cite{lim2017enhanced}, Set14 \cite{zeyde2012single}, CBSD68 \cite{martin2001database}, Flickr2K \cite{lim2017enhanced}, LSDIR \cite{li2023lsdir};
    \item MRI scans: FastMRI \cite{zbontar2018fastmri}, CMRxRecon \cite{wang2024cmrxrecon};
    \item Computed tomography scans: LIDC-IDRI \cite{armato2011lung};
    \item Fluorescence microscopy images: FMD \cite{zhang2019poisson}; 
    \item Real motion blur images: Kohler \cite{kohler2012recording};
    \item Multispectral satellite images: NBU \cite{meng2021pansharpening}.
\end{itemize}

\section{Evaluation}

Reconstruction methods can be evaluated on datasets with the method \code{Trainer.test} using metrics defined in our framework.
These are written in \code{deepinv} as \code{m = metric(x\_hat, x)} in the case of full-reference metrics, or as \code{m = metric(x\_hat)} for no-reference metrics, and provide common functionality such as input normalization and complex magnitude.

Following the distortion-perception trade-off in image reconstruction problems \cite{blau2018perception}, 
the library provides common popular distortion metrics such as PSNR, SSIM \cite{wang2004image}, and LPIPS \cite{zhang2018unreasonable}, 
as well as no-reference perceptual metrics such as NIQE \cite{mittal2012making} and QNR \cite{yeganeh2012objective}, which can be used to evaluate the quality of the reconstructions.

\section{Philosophy}

\subsection{Coding Practices}

DeepInverse is coded in modern Python following a test-driven development philosophy.
The code is unit-, integration- and performance-tested using \code{pytest} and verified using \code{codecov} (automatically via GitHub actions),
and is compliant with PEP8 using \code{black}.  To encourage reproducibility, the library passes random number generators
for all random generation functionality. Architecturally, \code{deepinv} is implemented using an object-oriented framework
where base classes provide abstract functionality and interfaces (such as \code{Physics} or \code{Metric}),
sub-classes provide specific implementations or special cases (such as \code{LinearPhysics}) along with methods inherited
from base classes (such as the operator pseudoinverse), and mixins provide specialized methods. This framework
reduces code duplication and makes it easy for researchers, engineers, and practitioners to implement new or specialized
functionality while inheriting existing methods.

\subsection{Documentation}

The library provides a \textbf{user guide}, quickstart and in-depth \textbf{examples} for all levels of user, and individual API documentation
for classes. The documentation is built using Sphinx. We use Sphinx-Gallery \cite{najera2023sphinxgallery} for generating Jupyter notebook demos, which 
are automatically tested and included in the documentation. The user guide provides an overview of the library and is intended as a starting point for new users, whereas the API lists all classes and functions, which are intended for advanced users. The user guide also serves as a computational imaging tutorial, providing an overview of
common imaging operators and reconstruction methods.
The documentation of most classes includes a usage example which is automatically tested using \code{doctest}, and 
a detailed mathematical description using latex with shared math symbols and notation across the whole documentation.

\section{Perspectives}

DeepInverse is a dynamic and evolving project and this paper is merely a snapshot of ongoing progress (release v0.3.0). The community is continuously contributing more methods, such as more realistic physics operators and more advanced training techniques, which reflect the state-of-the-art in imaging with deep learning, aiming to address the needs and interests of researchers and practitioners alike.

\section*{Acknowledgements}

J. Tachella acknowledges support by the French National Research Agency (Agence Nationale de la Recherche) grant UNLIP (ANR-23-CE23-0013) and the CNRS PNRIA deepinverse project.
M. Terris acknowledges support by the BrAIN grant (ANR-20-CHIA-0016).
F. Sarron, P. Weiss, M.H. Nguyen were supported by the ANR Micro-Blind ANR-21-CE48-0008.
Thomas Moreau was supported from a national grant attributed to the ExaDoST project of the NumPEx PEPR program, under the reference ANR-22-EXNU-0004.
J. Hertrich is supported by the German Research Foundation (DFG) with project number 530824055.
Z. Hu acknowledges funding from the Swiss National Science Foundation (grant PZ00P2\_216211). Thomas Davies is supported by UKRI EPSRC (grants EP/V006134/1 and EP/V006177/1).
S. Neumayer acknowledges funding from the German Research Foundation (DFG) with project number 543939932. 
We thank the BASP Laboratory at Heriot-Watt University (https://basp.site.hw.ac.uk/) for insightful discussions, and their contribution on the radio astronomy application.
The authors acknowledge the Jean-Zay high-performance computing center, using HPC resources from GENCI-IDRIS (Grants 2021-AD011012210, 2024-AD011015191).

\bibliographystyle{unsrt}
\bibliography{paper}

\end{document}